\documentclass[prl,twocolumn,amssymb]{revtex4}
\usepackage{graphicx}

\begin{document}

\title{Vortex Lattice in Planar Bose-Einstein Condensates with Dipolar Interactions}
\author{Jian Zhang and Hui Zhai}
\affiliation {Center for Advanced Study, Tsinghua
University, Beijing, 100084, P. R. China}
\date{\today}
\begin{abstract}

In this letter we investigate the effects of dipole-dipole
interactions on the vortex lattices in fast rotating Bose-Einstein
condensates. For single planar condensate, we show that the
triangular lattice structure will be unfavorable when the $s$-wave
interaction is attractive and exceeds a critical
value. It will first change to a square lattice, and then become more and more flat with the increase
of $s$-wave attraction, until the collapse of the condensate. For
an array of coupled planar condensates, we discuss how the
dipole-dipole interactions between neighboring condensates compete
with the quantum tunneling processes, which affects the relative
displacement of two neighboring vortex lattices and leads to the
loss of phase coherence between different condensates.

\end{abstract}
\maketitle

{\it Introduction.} During the latest ten years an extensive study
of the many-body physics of ultracold atomic gases
once again reveals that the interactions between particles play a
significant role in a many-body system, such as determining the
structure of a vortex lattice\cite{Ho,Baym,Cornell}, and
driving quantum phase transitions\cite{Zoller,Bloch}. Recently the
Bose-Einstein condensate (BEC) of chromium has been realized,
where the magnetic dipole-dipole interaction is $36$ times larger
than the alkali atoms\cite{pfau}; and it is very hopeful that the
quantum degeneracy of polar molecules, which have large electric
dipole moments, may also be achieved soon\cite{Marcassa}. In the
study of these systems, it is necessary to incorporate the dipolar
interactions between atoms (molecules). Theoretical works have revealed
many distinctive effects caused by the dipolar interactions. For
examples, owing to its anisotropic nature, the stability of a
dipolar BEC depends on the trap geometry\cite{You,Odell} and its
excitation spectrum exhibits roton-maxon structure\cite{roton};
and owing to its long-range nature, the dipolar Bosons in optical
lattice provide supersolid and checkerboard phases because of the
non-negligible interactions between nearest-neighboring
sites\cite{supersolid}.

\begin{figure}[bp]
\begin{center}
\includegraphics[width=8.5cm]
{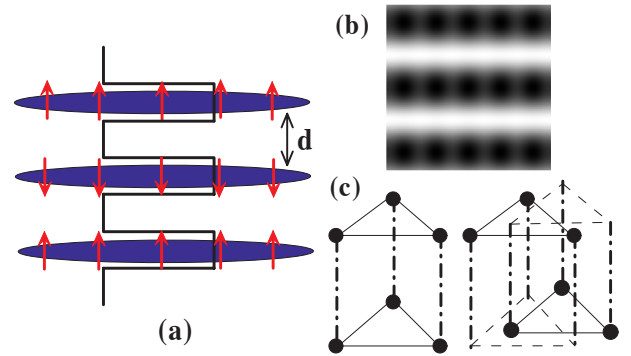}\caption{(color online) (a) A schematic of the model.
The blank line stands for the one-dimensional optical lattice, and the red
arrows denote the directions of spin polarization. (b) The density profile of a flat vortex lattice, the blank regime is the low density regime and the bright regime is the high density regime. (c) A schematic
of the coincident phase (left) and the staggered phase
(right).\label{model}}
\end{center}
\end{figure}

In this letter we are interested in how these characteristics of
the dipolar interactions manifest themselves in the structure of
vortex lattice in a fast rotating BEC. We consider a BEC of chromium which is fast
rotated with respect to $z$ direction and loaded into a
one-dimensional lattice imposed along the $z$ direction, as schematically illustrated in Fig.(\ref{model}a).  In each site, it is a planar
condensate containing a large number of vortices. When the sites are widely separated, these condensates are isolated from each other, and when the sites
get closer, they are coupled with their neighborhoods via quantum
tunnelling and the long-range dipolar interactions. Here we
consider the case that the spins of atoms are all
polarized and the spin polarizations
are antiferromagnetically arranged among the lattice
sites, therefore both the intra-layer and the inter-layer dipolar
interactions are repulsive. This antiferromagnetic arrangement can be realized by manipulating the configuration of external magnetic field, for example, by using the magnetic field nearby permanent-magnet material\cite{magnetic}.

This work contains two parts. First, we consider an isolated
planar condensate under fast rotation. When the rotation frequency is very close to the trapping frequency, the size of vortex core becomes comparable to the spacing between vortices, and its effect becomes important\cite{Ho,Cornell}. Therefore the repulsive $s$-wave interaction
favors a triangular vortex lattice where the vortex cores are
arranged in a most symmetric way so that the modulation of the condensate density is spatially homogenous, while the
attractive interaction favors the lattice structure such as the square lattice and the flat lattice, which are less
symmetric than the triangular lattice. However, in the absence of
the dipolar interactions, such lattices can not exist
because the condensate itself collapses when the $s$-wave interaction
becomes attractive. The presence of dipolar interaction prevents
the condensate from collapse for small $s$-wave attraction and
thus the effect of attractive interaction on the lattice structure
becomes observable. Here we are going to show that a triangular
lattice will be unfavorable in the parameter region that the
strength of $s$-wave attraction exceeds a critical value but is
not enough to induce collapse. With the increase of $s$-wave
attraction, the vortex lattice will experience a structure
transition from triangular to square, and then to very flat lattice
as shown in Fig. \ref{model}(b).

Secondly, we study the effects of the coupling between the neighboring
condensates. In the situation that the quantum tunneling is
dominative, the condensates in different sites are coherent, and
the positions of vortices in different layers are coincident with
each other (see the left one of Fig.\ref{model}(c)). However, because
the inter-layer dipolar interacting energy will be reduced when
the high density region of one condensate coincides with the low
density region of its neighboring condensate, it favors that the
vortex lattices are stagger-arranged among the lattices(see
the right one of Fig.\ref{model}(c)). Here we will show that the
competition between the quantum tunneling and the inter-layer
dipolar interactions will lead to a quantum phase transition from
the coincident phase to the staggered phase. Moreover, the tunneling amplitude will be suppressed
exponentially when the vortex lattices are staggered, and
consequently the fluctuation of relative phases between
neighboring condensates will be enhanced. Therefore, this quantum
phase transition is characterized by the loss of phase coherence
between sites, in analogy with the superfluid-Mott insulator
transition.

{\it Model and Method.} We consider a two-dimensional BEC, which
is trapped by a harmonic potential $m\omega^2 r^2/2$ in the $xy$ plane
and fast rotated with fixed frequency $\Omega$, the system is
described by following energy functional
\begin{equation}
K=\int d^2{\bf r}\Psi^*({\bf r})\hat{H}_{0}\Psi({\bf r})+\int
d^2{\bf r}d^2{\bf r^\prime}\rho({\bf r})V({\bf r}-{\bf
r^\prime})\rho({\bf r^\prime}),
\end{equation}
where $\Psi({\bf r})$ is the macroscopic wave function of the
condensate, $\rho({\bf r})$ is the condensate density, and $V({\bf
r}-{\bf r^\prime})$ is the interactions between atoms which
include both the $s$-wave contact interaction and the magnetic
dipolar interaction. The single particle Hamiltonian $\hat{H}_{0}$
is written as $-\hbar^2\nabla^2/(2m)+m\omega^2 r^2/2-\Omega L$,
whose eigenstates are Landau levels in the fast rotating limit
where $\Omega$ is very close to $\omega$, and the interaction energy is much weaker than the spacing between different Landau levels $2\hbar\Omega$ when the condensate density is very low, therefore we can restrict
ourselves in the lowest Landau level(LLL)\cite{Ho}. When the vortices form
a uniform lattice, the wave function $\Psi({\bf r})$, excluding
the Gaussian factor, is a double-periodic analytical function, and
can be uniquely given by the Jacobi theta
function\cite{Ho-stripe}. Although the
Gaussian profile of the global density is unstable with respect to very
weak lattice distortion\cite{Baym}, the LLL wave function serves
very well as a trial wave function to determine the vortex
lattice structure. For examples, this wave
function has been successfully used to explain the dynamic
formation of vortex stripe observed in a recent experiment
\cite{Cornell-stripe,Ho-stripe}; and to predict the structure
transition of vortex lattice from triangular to square in a
two-component condensate\cite{Ho-stripe}, which has
been observed later\cite{Cornell-twocom}.

By denoting ${\bf b}_{1}$ and ${\bf b}_{2}$ as the basis vectors of the
lattice, a uniform vortex lattice in the LLL is characterized by two parameters, which 
are ${\bf b}_{2}/{\bf b}_{1}=u+iv$ describing the lattice type and
$v_{c}=b_{1}^2v$ denoting the area of a unit cell. The condensate density takes the form as
$\rho({\bf r})={1}/{(\pi\sigma^2)}\sum_{{\bf K}}\tilde{g}_{{\bf
K}}\exp(i{\bf K}\cdot{\bf
r})\exp(-r^2/\sigma^2)$\cite{Ho,Ho-stripe},
where the summation is taken over all reciprocal lattice vectors
$m_{1}{\bf K}_{1}+m_{2}{\bf K}_{2}$, with ${\bf K}_{1}$ and ${\bf
K}_{2}$ being the basis of the reciprocal lattice. Here $\sigma$ is the condensate radius which is given by $\sqrt{[a_{\perp}^{-2}-\pi v_{c}^{-1}]^{-1}}$ where $a_{\perp}=\sqrt{\hbar/(m\omega)}$. The coefficients $\tilde{g}_{{\bf K}}$ denotes $g_{{\bf K}}/g_{0}$, and $g_{{\bf K}}$ is
explicitly written as
$g_{{\bf
K}}=(-1)^{m_1+m_{2}+m_{1}m_{2}}e^{-v_{c}|K|^2/8\pi}\sqrt{v_{c}/2}$,
with $v_{c}|K|^2=(2\pi)^2v^{-1}[(vm_{1})^2+(m_{2}+um_{1})^2]$.

In the LLL region the mean value of the single particle Hamiltonian
$E_{0}$ turns out to be
$\hbar(\omega_{\perp}-\Omega)\sigma^2$\cite{Ho-stripe},
which is independent of the vortex lattice structure. The energy of $s$-wave interaction,
$V_{\text{s}}({\bf r}-{\bf r^\prime})=g\delta({\bf r}-{\bf
r^\prime})$, can also be evaluated as $
E_{\text{s}}=g\int d^2\mathbf{r}\rho^2({\bf r}) \simeq
gI/(\pi\sigma^2)
$\cite{Ho-stripe},
where $I$ denotes the summation of $\sum_{{\bf K}}|g_{{\bf K}}/g_{0}|^2$.
When the $s$-wave interaction is repulsive, i.e. $g>0$, we can
find that the triangular lattice is stable by minimizing $E_{\text{s}}$, and the minimization of the total energy,
including $E_{0}$ and $E_{\text{s}}$, yields a finite value of
condensate radius $\sigma$. However, when $g$ is negative, the
minimization of the total energy results in $\sigma$ vanishing, which
implies the collapse of the condensate.

{\it Single Planar Condensate.} Within the LLL mean field ansatz, we
are going to investigate how the dipolar interactions affect the
structure of vortex lattice of a planar condensate. Considering a very deep lattice, at each lattice site the motion of atoms along $z$ direction is strongly confined, and the magnetic dipole moment ${\bf \mu}$ of atoms are also
polarized along $z$ axis, the magnetic dipolar interaction in $xy$ plane reads
\begin{eqnarray}
V_{\text{d}}(\mathbf{r}_1,\mathbf{r}_2)
=\frac{\mu_0\mu^2}{4\pi}\frac{1}{|\mathbf{r}_1-\mathbf{r}_2|^3}\label{twoDdipolar},
\end{eqnarray}
and the dipolar interaction energy $E_{\text{d}} $ is given by
$\int d^2 {\bf r_1} d^2{\bf r_2} \rho({\bf r_1})V_{\text{d}}({\bf
r_1}-{\bf r_2})\rho({\bf r_2}) $. Denoting the relative
displacement ${\bf r}={\bf r}_1-{\bf r}_2$, and the center of mass
displacement ${\bf R}=({\bf r}_1+{\bf r}_2)/2$, we can first
integrate ${\bf R}$ out and $E_{\text{d}}$ turns out to be
\begin{eqnarray}
E_{\text{d}} =\frac{\mu_0\mu^2}{8\pi^2\sigma^2}\sum_{{\bf
K}}\left(\left|\frac{g_{{\bf K}}}{g_{0}}\right|^2\int d^2{\bf
r}\frac{1}{r^3}e^{i{\bf K}\cdot{\bf
r}}e^{-r^2/(2\sigma^2)}\right).\label{d-int}
\end{eqnarray}
This integral equals to $2\pi\int dr
J_0(Kr)e^{-r^2/(2\sigma^2)}/r^2$, which has shot-range divergence and therefore requires a short-range cut-off $\Lambda$. Here $\Lambda$ should be choose as $4a_{z}$ where $a_{z}$ is the longitudinal harmonic length of each lattice site\cite{Lambda}.   It is important to notice
that a dimensionless value $W$ defined as bellow is regular, i.e.
\begin{equation}
W=\sum\limits_{{\bf K}}\left\{\left|\frac{g_{{\bf
K}}}{g_{0}}\right|^2\int
d\tilde{r}\frac{1}{\tilde{r}^2}\left[J_0(K\tilde{r}a_{\perp})e^{-\tilde{r}^2a_{\perp}^2/(2\sigma^2)}-1\right]\right\}
\end{equation}
with $\tilde{r}=r/a_{\perp}$, and the remainder term depending on
$\Lambda$ is proportional to $I$, thus the total dipolar energy of
a rotating condensate can be viewed as an isotropic and locally
repulsive part plus an attractive part, because $W$ is always
negative. The total interaction energy including the dipolar and
s-wave interaction turns out to be
\begin{eqnarray}
E_{\text
{int}}&=&\frac{1}{\pi\sigma^2}\left[\left(\frac{\mu_0\mu^2}{4\Lambda}+g\right)I
+\frac{\mu_0\mu^2}{4a_{\perp}}W\right]\nonumber\\
&=&\frac{\mu_{0}\mu^2}{4\pi\sigma^2 a_{\perp}}(\alpha I+W),
\end{eqnarray}
where $\alpha$
denotes $a_{\perp}/\Lambda+4g a_{\perp}/(\mu_{0}\mu^2)$ which can
be gradually changed in a wide range because the $s$-wave
interaction constant $g$ can be tuned via Feshbach resonance
technique\cite{Pfau-resonance}. The structure of vortex lattice in
equilibrium can be obtained by minimizing $E_{\text{int}}$, and the
collapse of condensate occurs when $E_{\text{int}}$ becomes negative.

\begin{figure}[tbp]
\begin{center}
\includegraphics[width=9.0cm]
{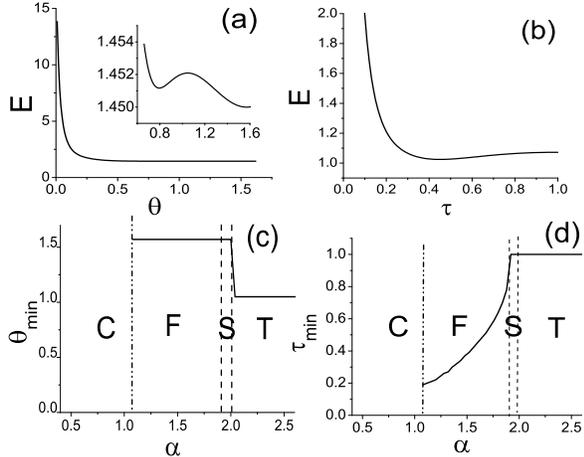}\caption{(a): The dimensionless value of $E$, which is defined as $\alpha I+W$, as a function of
the lattice angle $\theta$ with $\alpha=1.92$, where
$\theta=\arccos{({\bf b}_1\cdot{\bf b}_2)}/(b_1b_2)$. The insert is a zoom-in plot of the range $\theta\in[\pi/3,\pi/2]$. (b):  $E$ as a function of $\tau=|{\bf b}_{2}/{\bf b}_{1}|$ with $\alpha=1.60$. 
(c) the lattice angle of the ground state $\theta_{\text{min}}$ as a function of parameter $\alpha$ and (d) the radio $\tau$ of the ground state as a function of $\alpha$. Here  {\textbf C} denotes collapse, {\textbf F} denotes flat lattice, $\textbf{S}$ denotes square lattice and {\textbf T} denoting triangular lattice.\label{IWcompare}}
\end{center}
\end{figure}

The results of the minimization of the mean field energy are shown in Fig.(\ref{IWcompare}). In the regime that the repulsive interaction is dominant, the ground state is a triangular lattice with $\theta_{\text{min}}=\pi/3$ and $|{\bf{ b}}_{2}/{\bf b}_{1}|=1$. When $\alpha$ decreases below $2.0$, the triangular lattice becomes unstable and will be replaced by the square lattice, which can be seen from Fig.(\ref{IWcompare}a) where the energy at $\theta=\pi/2$ is lower than $\theta=\pi/3$. Furthermore, the energy takes its minimum at the place where $|{\bf b}_{2}/{\bf b}_{1}|$ is smaller than unity when $\alpha$ is smaller than $1.9$, as shown in Fig.(\ref{IWcompare}b). From Fig.(\ref{IWcompare}c)and (\ref{IWcompare}d) one can see that the vortex lattice changes from a triangular one to a square one, and then becomes more and more flat until the collapse of the whole condensate, as the $s$-wave attraction increases. In this calculation we take $\Lambda/a_{\perp}=0.1$ which is a typical values of current experiments, for chromium the square lattice occurs when $a_{\text{s}}$ is between $-2.32a_{0}$ and $-2.28a_{0}$, and the flat lattice occurs when $a_{\text{s}}$ is between $-2.55a_{0}$ and $-2.32a_{0}$, where $a_{0}=0.0529\text{nm}$.

{\it Coupled Planar Condensates}. Now we are going to study the
effects of coupling between nest-neighboring condensates. Provided that the energy scale of the
inter-layer coupling processes is much smaller than the
intra-layer energy scale, the structure of vortex lattice will not be affected by the
inter-layer coupling.
Here we are interested in how the inter-layer dipolar
interactions affect the relative displacement of vortex lattices in
different layers, and therefore we focus on a typical case of
triangular vortex lattice. 
Neglecting the $s$-wave interaction
between layers, the inter-layer interaction is 
\begin{equation}
V^\prime_{\text{d}}({\bf r}_{1},{\bf r}_{2})
=\frac{\mu_0\mu^2}{4\pi}\frac{2d^2-r^2}{(d^2+r^2)^{5/2}}.
\end{equation}
where $d$ is the distance between two layers, and
$\mathbf{r}=\mathbf{r}_1-\mathbf{r}_2$ denotes the relative
displacement in the $xy$ plane. In the large vortex number limit
the inter-layer interaction energy turns out to be
\begin{equation}
E^\prime_{\text{d}}=\frac{\mu_0\mu^2}{4\pi\sigma^2 d}\sum_{{\bf
K}}\left(\left|\frac{g_{{\bf K}}}{g_{0}}\right|^2
e^{i\mathbf{K}\cdot\bf{r}_0}F(|{\bf K}|,d)\right) \label{inter-layerintera}
\end{equation}
where $F(|{\bf K}|,d)$ denotes a dimensionless integral 
\begin{equation}
F(|{\bf K}|,d)=
\int d\tilde{r}J_0(|{\bf K}|\tilde{r}d)e^{-\tilde{r}^2d^2/(2\sigma^2)}\frac{\tilde{r}(2-\tilde{r}^2)}{(1+\tilde{r}^2)^{5/2}}
\end{equation}
with $\tilde{r}=r/d$, and $\mathbf{r}_0$ is the relative displacement between two
neighboring vortex lattices. Because $|g_\mathbf{K}|$
exponentially decreases as the increase of $|\mathbf{K}|^2$, it is
sufficient to keep the terms with small value of $|K|$ in the
summation, namely, $K_{0}=0$ and
$K_{1}=|\mathbf{K}_{\pm1,0}|=|\mathbf{K}_{0,\pm1}|=|\mathbf{K}_{\pm1,\mp1}|$
for the triangular lattice. Thus
\begin{eqnarray}
E^\prime_{\text{d}}=\frac{\mu_0\mu^2}{4\pi\sigma^2d}
\left(F(K_{0},d)+2CF(K_{1},d)\sum
^\prime\cos(\mathbf{K}\cdot\mathbf{r}_0)\right).
\end{eqnarray}
Here both $F(K_0, d)$ and $F(K_1, d)$  are positive, the constant $C$ denotes the value of
$|g_{\mathbf{K}_{1,0}}/g_0|^2$ ($=|g_{{\bf
K}_{0,1}}/g_{0}|^2=|g_{{\bf K}_{1,-1}}/g_{0}|^2$ ) which equals
to $0.0532$ for the triangular lattice, and $\sum^\prime$ denotes
the summation over $\mathbf{K}_{1,0}$,$\mathbf{K}_{0,1}$ and
$\mathbf{K}_{1,-1}$. Taking the symmetry of the triangular lattice
into account, we set ${\bf r}_0=x({\bf b}_{1}+{\bf b}_{2})$ with $x\in[0,1/2]$,
$E^\prime_{\text{d}}$ takes its minimum at $x=1/3$, which means
that the vortex cores of one lattice displace to the
center of unity cells of its neighboring lattice.

\begin{figure}[tbp]
\begin{center}
\includegraphics[width=8.0cm]
{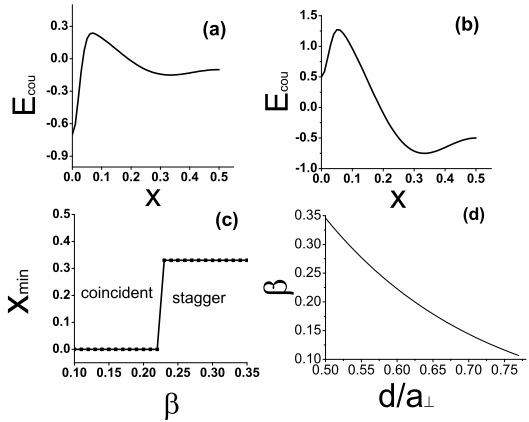}\caption{ (a,b) The energy $E_{\text{cou}}$ (in the unit of $t/2$) as a
function of ${\bf r}_{0}=x({\bf b}_{1}+{\bf b}_{2})$, for two
typical parameters of $\beta$. $\beta=0.1$ and
$E_{\text{cou}}$ takes its minimum at $x=0$ for (a); $\beta=0.5$
and $E_{\text{cou}}$ takes it minimum at $x=1/3$ for (b). (c) The value of
$x_{\text{min}}$, which denotes the value where $E_{\text{cou}}$ takes its minimum, as a function of $\beta$.
(d) $\beta$ as a function of lattice spacing $d$, with $t$ fixed at $\mu_{0}\mu/(10\pi\sigma^2 a_{\perp})$. \label{stagger}}
\end{center}
\end{figure}

Another inter-layer process is the quantum tunneling between
neighboring condensates. Within the LLL mean field theory,
the tunnelling energy can be expressed as
$
E_{\text{t}}=-t\cos\phi\exp(-N_{\text{v}}{\pi}r^2_0/v_{\text{c}})/2
$
where $N_{\text{v}}$ is the number of vortices, $t$ is
the hopping amplitude between two neighboring sites, and $\phi$ is the relative phase between two neighboring condensates. This term favors $\bf{r}_0=0$, which corresponds
to binding the vortices in different layers at same positions in the $xy$
plane\cite{zhai,Ueda}. Therefore the total energy of the processes
coupling two neighboring condensates, $E_{\text{cou}}$,
includes $E^\prime_{\text{d}}$ and $E_{\text{t}}$. By minimizing
$E_{\text{cou}}$, it can be found that the competition between these
two processes will result in a transition of the ground state. Two typical curves of
$E_{\text{cou}}$ are shown in Fig.\ref{stagger}(a) and (b),  in the case (a) the vortex lattices are coincident and in the case (b)
they are staggered. As illustrated in Fig.\ref{stagger}(c),
this transition is driven by the parameter $\beta$ defined as
${\mu_0\mu^2CF(K_{1},d)}/{(t\pi \sigma^2d)}$, and it
is very a sharp one. Experimentally, $\beta$
can be gradually increased by decreasing the
lattice spacing $d$, with  the barrier height simultaneously increased to keep $t$ unchanged. As illustrated in Fig.\ref{stagger}(d), a transition to staggered phase occurs when $d/a_{\perp}$ is decreased below $0.6$.

It is also worthwhile to consider the relative phase fluctuations
between two neighboring condensates. Because $\delta\phi$ is
proportional to $1/\sqrt[4]{t\exp(-N_v\pi r_0^2/v_{\text{c}})}$, the
denominator of which will be suppressed to a exponentially small value
when $r^2_{0}$ becomes comparable to $v_{\text{c}}$ in the staggered
phase, the transition of ${\bf r}_{0}$ is accompanied by the loss
of phase coherence between different layers and the
suppression of particle number fluctuations in each site. In this
sense, this transition is similar to the superfluid-Mott insulator
transition. However, a peculiar point is that this transition is
driven by nearest-neighboring interactions instead of the on-site
interactions.

The authors would like to thank Professor C. N. Yang for
encouragement and valuable support. We thank Professor T. L. Ho
for his valuable comments, and L. Chang and Q. Zhou for helpful
discussions. We 
thank N. Cooper for helpful correspondence. This work is supported by NSF China (No. 10247002 and
10404015).

{\it Note added.} After finishing an initial version of this paper, we became aware of the work by
Cooper, Rezayi and 
Simon in which the vortex lattice stucture of a planar dipolar condensate was
also discussed\cite{Cooper}.


\begin{thebibliography}{99}
\bibitem{Lambda}
We get the dipolar interaction Eq.(\ref{twoDdipolar}) by integraling over the $z$ direction, i.e. 
$\int dze^{-2z^2/a^2_{\text{z}}}(2z^2-r^2)/[a_{\text{z}}(z^2+r^2)^{5/2}]$, and then take limit of small $a_{\text{z}}$. For $r>4a_{\text{z}}$, it turns out to be Eq.(\ref{twoDdipolar}), and for $r<4a_{\text{z}}$ it vanishes as $a_{\text{z}}$ goes to zero.
\end{thebibliography}
\end{document}